\documentclass[12pt]{article}
\usepackage{amssymb}
\def\C{\mathbb{C}}
\def\R{\mathbb{R}}
\def\ap{{\alpha '}}
\def\B{\Theta}
\makeatletter

\renewcommand\section{\@startsection{section}{1}{\z@}%
                                     {-6.25ex\@plus -1ex \@minus -.2ex}%
                                     {1.5ex \@plus .2ex}%
                                     {\reset@font\large\bfseries}}
\renewcommand\subsection{\@startsection{subsection}{2}{\z@}%
                                     {3.25ex \@plus1ex \@minus.2ex}%
                                     {-1em}%
                                     {\reset@font\normalsize\bfseries}}
\@addtoreset{equation}{section}
\makeatother

                       \def\pl{\partial}
\def\be{\begin{equation}}            \def\ba{\begin{eqnarray}}
\def\ee{\end{equation}}                    \def\ea{\end{eqnarray}}
\newcommand{\ar}{\begin{array}}            \newcommand{\er}{\end{array}}
\def\nn{\nonumber}                         

        \def\k{{\cal k}}

        \def\B{{\cal B}}      

         \def\e{\epsilon }      
           \def\k{\kappa }

\def\bz{{\bar z}}
\def\i{{\rm i}}
\def\Im{{\rm Im}}
\def\bJ{{\bar J}}

\def\cH{{\cal H}}

\def\k{{\rm k}}
\def\bw{{\bar w}}
\def\bB{{\rm B}}
\def\B{{\Theta}}
\begin{document}
\thispagestyle{empty}
\setcounter{page}{0} 
\def\thefootnote{\fnsymbol{footnote}}
\begin{flushright}
  REPORT No.\ 13 1998/99 \\
  DESY 99-039 
\end{flushright} \vskip 2.0cm
\begin{center}\LARGE
{\bf D-branes and Deformation Quantization}
\end{center} \vskip 1.0cm
\begin{center}\large
           Volker Schomerus
 \footnote{E-mail  address: {\tt vschomer@x4u2.desy.de}}
\end{center}
\vskip0.5cm
\begin{center}
II. Institut f\"ur Theoretische Physik\\
Universit\"at Hamburg, D-22761 Hamburg, Germany\\[3mm]
and \\[3mm] 
Institut Mittag-Leffler, Royal Swedish Academy of Sciences\\
Aurav\"agen 17, S-182 62 Djursholm, Sweden  
\end{center}
\begin{center}
March 21, 1999
\end{center}
\vskip 1cm
\begin{abstract}
In this note we explain how world-volume geometries of D-branes can be 
reconstructed within the microscopic framework where D-branes are 
described through boundary conformal field theory. We extract  
the (non-commutative) world-volume algebras from the 
operator product expansions of open string vertex operators. For branes in 
a flat background with constant non-vanishing B-field, the operator products 
are computed perturbatively to all orders in the field strength. The resulting 
series coincides with Kontsevich's presentation of the Moyal product. 
After extending these considerations to fermionic fields we conclude 
with some remarks on the generalization of our approach to curved 
backgrounds.
\end{abstract}
\vfill
\setcounter{footnote}{0}
\def\thefootnote{\arabic{footnote}}
\newpage

\section{Introduction} 

It was observed many years ago that low energy effective 
actions of (super-)string theories possess solitonic solutions.
They are known as solitonic $p$-branes because of their localization 
along certain $p+1$-dimensional surfaces in the string-background. 
Later, Polchinski discovered a remarkable correspondence between 
such solitonic $p$-branes and D$p$-branes, i.e.\ boundary conditions 
for open strings (for a review see \cite{Pol}). This makes it possible 
to study branes through the `microscopic' techniques of boundary conformal 
field theory. 
\smallskip

In the microscopic approach, D-branes are characterized by 
the way in which they couple to closed string states, i.e. by the 
1-point functions of closed string vertex operators. An immediate 
question is how these couplings relate to the hyper-surfaces
one meets in the macroscopic description of D-branes. Here, 
we shall address a systematic reconstruction of  D-brane world-%
volumes from their world-sheet description. We argue that the 
information on the geometry of the world-volumes is encoded 
in the operator products of open string vertex operators. The 
idea to retrieve geometrical data from operator products is 
not new, but so far it was mainly applied to closed string 
vertex operators (see e.g.\ \cite{SNCG,LLS}). 
\smallskip

When D-branes are placed in a background which carries a non-vanishing 
B-field, our procedure will lead us to a deformation of the algebra of 
functions on the classical world-volume. This phenomenon was first observed 
by Douglas and Hull \cite{DoHu} in the example of a 2-dimensional brane 
wrapping a 2-torus (see also \cite{NCG,LLS,ChHo} for more 
recent developments in this direction) and it is obviously related to 
the structure of compactifications of M(atrix) theory on non-commutative 
tori \cite{CDS}. Let us remark that non-commutativity enters quite 
naturally in a theory of open strings. In fact, open string vertex 
operators are inserted along a one dimensional line (the boundary of the 
world-sheet) so that their insertion points are canonically ordered. 
The product of two open string vertex operators usually depends
on the order in which they are inserted and hence it provides an 
obvious `source' for non-commutative geometries. 
\smallskip

In this short note we restrict our presentation to the case of 
constant B-fields on a flat background. We shall use 
standard techniques from conformal perturbation theory (see e.g.\ 
\cite{CFT}) to derive an explicit formula for the non-commutative 
multiplication in the world-volume algebra. It will turn out as 
the Moyal deformation of the classical algebra of functions on 
the world-volume. The deformation depends on the string tension 
and on the B-field which enters through the anti-symmetric tensor 
$B (1+B^2)^{-1}$ (see eq.\ (\ref{ast1}) below), in agreement with 
the recent analysis of Chu and Ho \cite{ChHo}. At the end of the text, 
we extend these considerations to the fermionic sector (see eq.\ 
(\ref{Cdef})).  
\smallskip

While some of the formulas below are not new (see e.g. \cite{DoHu,NCG}), 
our techniques are designed for generalizations to non-trivial 
backgrounds and, in particular, to perturbative studies in the 
framework of non-linear $\sigma$-models where the B-field is then 
allowed to depend on coordinates. The approach also displays clearly 
the remarkable relation between open string theories and quantization. 
In the context of topological open strings this relation is beautifully 
illustrated by the recent work of Cattaneo and Felder \cite{CaFe}  
on Kontsevich's quantization formula \cite{Kon}. Here, we shall 
see the background metric entering the deformation and the results 
of topological theories appear only in a very special limit.

\section{Open Strings and D-branes in a Flat Background} 

To set our stage, we briefly consider the standard bosonic open string 
moving on a flat $d$-dimensional Euclidean background, i.e.\ on the space 
$\R^d$. Its world-sheet description involves $d$ free bosonic fields 
$X^i(z,\bz), i= 1,\dots,d,$ living on the upper half $\Im z \geq 0$ of 
the complex plane with Neumann boundary conditions imposed along the 
boundary $\Im z = 0$. The explicit construction of this field theory 
is encoded in the formula
\be X^i(z, \bz)  \ = \ \hat x^i - \i\, \frac{\ap}{2}\, \hat p^i \, \ln
   z\bz + \i\, \sqrt{\frac{\ap}{2}} \, \sum_{n\neq 0} \, \frac{a^i_n}{n}\,  
\left(
   z^{-n} + \bz^{-n} \right) \ \ , \label{Xcon}\ee
where $a^i_n, n < 0 \ (n > 0),$ create (annihilate) oscillations
of the open string and $\hat x^i, \hat p^i$ describe the string's
center-of-mass coordinate and momentum. They obey the usual
commutation relations involving some constant metric $G_{ij}$ 
on the background. The parameter $\ap$ is the inverse of the
string tension, up to some normalization. 
\smallskip

{}From the free bosonic fields $X^i(z,\bz)$ on the upper half plane
we can build many new fields, in particular the chiral currents
$J^i(z), \bJ^i(\bz) $ and open string vertex operators $V_\k(x)$, 
\ba J^i(z) \ = \ 2\i \, \partial X^i(z, \bar z) \ \ & , &  \ \
   \bJ^i(\bz) \ = \ 2\i\, \partial X^i(z, \bz) \ \ ,\nn \\[2mm] 
   V_\k(x) \ = \ :e^{\i \, \k_i X^i(x)} : \ \ \ \ \ \  \mbox{ where }  & & 
   X^i(x) \ = \   \hat x^i - \i\, \ap \, \hat p^i\,  \ln x
   + \i \, \sqrt{2 \ap} \sum_{n\neq 0}\,  \frac{a^i_n}{n}\,  x^{-n} \ \ \nn
\ea 
is the boundary value $X^i(x), x \in \R,$ of the bosonic field
$X^i(z,\bz)$. The open string vertex operators satisfy the
following elementary operator product expansions (see e.g.
\cite{CFT}) :
\ba
V_{\k}(x_1)\, V_{\k'}(x_2)  \ \sim \
  \frac{1}{(x_1-x_2)^{\ap\k \k'}}
  & &\hspace*{-7mm} V_{\k + \k'} (x_2) \ + \ \dots\
  \ \mbox{ for }\ \ \
   x_1 > x_2 \ \ , \label{OPE} \\[2mm]
 J^i(z) \, V_\k(x) \ \sim \  \frac{2 \ap \k_i}{z-x} \ V_\k(x)    
 \ + \ \dots \ \ & , & \ \   
\bJ^i(\bz) \, V_\k(x) \ \sim \ \frac{2 \ap \k_i}{\bz -x} \ V_\k(x)
 \ + \ \dots \ \  . \nn
\ea
Throughout the text we shall raise and lower indices with the 
help of the metric $G$ and the product $k k'$ is defined through 
$k k' = k^i k'_i$. In all three relations we displayed only the 
most singular term and represented the rest of the expansion by 
dots. As we remarked above, operator products of open string 
vertex operators depend on the ordering of the insertion points, 
in general. This is why we specified the order $x_1 > x_2$ in the 
first equation, even though we are presently dealing with a
very special situation in which the relation for the
other order $x_1 < x_2$ happens to be of the same form.     
\smallskip

There exists a more elegant formulation of eqs.\ (\ref{OPE})
that makes use of the formal object 
$$ f(X(x)) \ = \ V[f ](x) \ :=\ \frac{1}{(2\pi)^{d/2}}\int_{\R^d}
                  d^d \k\,  \hat f (\k)\, 
                  V_\k(x) \ \ . $$
Here, $\hat f(\k)$ denotes the usual Fourier transform of
the function $f: \R^d \rightarrow \C$. The operator product
expansions (\ref{OPE}) become
\def\fri{\frac{1}{\i}}
\ba
V[f\, ](1)\, V[g ](0) & \sim & 
  \ V[f \, g\, ] (0) \ + \ \dots\ \ , \nn \\[2mm]
J^i(z) \, V[f\, ](x) & \sim & \fri \frac{2 \ap G^{ij}}{z-x} 
   \ V[\pl_j f\, ] (x)    
 \ + \ \dots \ \ , \label{OPE'} \\[2mm]
\bJ^i(\bz) \, V[f\, ](x) & \sim & \fri\frac{2 \ap G^{ij}}{\bz -x} \
V[\pl_j f\, ](x) \ + \ \dots 
 \ \ . \nn\ea
Note that in the first equation we have specialized to $x_1 = 1,
x_2 = 0$. The more general formula in rel.\ (\ref{OPE}) can be
recovered with the help of conformal transformations so that
we did not lose information in passing from eqs.\ (\ref{OPE})
to eqs.\ (\ref{OPE'}).

The first formula in rel.\ (\ref{OPE'}) is actually quite
remarkable since it describes the operator product expansion
of open string vertex operators in terms of a very simple
algebraic operation, namely in terms of the pointwise
multiplication of functions on $\R^d$.
\bigskip

Let us now change the background of our bosonic open string theory 
by switching on a B-field, i.e.\ by adding the following 
term to the action
\be S_B \ = \ \frac{1}{4 \pi \ap } \int_\cH dzd\bz
             \, B_{ij} \, J^i(z)\,  \bJ^j(\bz) \ \ ,
\label{SB} \ee
where $B_{ij}$ is some (constant) antisymmetric $d\times d$ matrix and 
the integral is performed over the upper half plane $\cH$.
Our main aim is to determine the influence $S_B$  on the operator 
product of open string vertex operators. At the moment, we shall 
only look for a suitable formulation of this problem. An explicit 
solution is then derived in the next section.  
\medskip

It is well known that the only effect of the term (\ref{SB})
is to modify the boundary condition along the real line so
that the bosonic fields satisfy 
$$ \pl_y X^i(z,\bz) \ = \ 
   B^i_{j} \, \pl_x X^j(z,\bz) \ \ \mbox{ where }
   \ \  z = x + \i y   \ \  $$
all along the boundary $z = \bz$. The usual
Neumann boundary condition $\pl_y X^i = 0$ is
recovered in the limit $B_{ij} \rightarrow 0$. We could go
ahead now and construct the new theory explicitly much as
we did this for Neumann boundary conditions above (see 
also \cite{ACNY},\cite{ChHo}). Of course,
the formulas for the fields $X^i(z,\bz)$ and their boundary
values $X^i(x)$ change and so does the operator product
expansion of the open string vertex operators, i.e.\ the
latter is no longer described by pointwise multiplication
of functions on the background. However, we may still think
about the new operator product expansions in terms of functions
on $\R^d$, if we are prepared to `deform' the pointwise
multiplication. In other words, we define a new product
$(f,g) \mapsto f \star g$ for functions $f,g: \R^d
\rightarrow \R$ by   
\be \left( V[f\, ](1) \, V[g ](0)\right)^B \ \sim \
   V[f\star g\, ] (0) \ + \ \dots \ \ . 
\   \label{ast0}
\ee
The superscript $B$ was placed on the left hand side to
indicate that the operator product is to be taken in a
theory in which the B-field is turned on. The
multiplication $\star$ on the right hand side depends
on $B_{ij}$ and on the parameter $\ap$. 

Even before having constructed $\star$, we can predict
some of its properties. The new multiplication will not
be commutative, in general. On the other hand, from the
sewing constraints of open string theory (see \cite{Lew},
\cite{sew},\cite{Run}) and the triviality of the fusing matrix in a theory
of free bosonic fields it is possible to deduce that {\it $\star$ is 
necessarily associative}. Consequently, when functions $f:\R^d \mapsto
\R$ are equipped with the product $\star$ they generate a
(non-commutative) associative deformation of the algebra
of functions on the background. This algebra depends on
the anti-symmetric matrix $B_{ij}$ and on the parameter $\ap$. 
\medskip

Open strings with Neumann boundary conditions can be interpreted 
in terms of a D-brane that fills the whole background. We can 
certainly generalize our considerations to D-branes of smaller 
dimension. In this case one has to impose Dirichlet boundary 
conditions in some directions and only those open string vertex 
operators survive that carry momentum tangential to the brane's 
world-volume. All constructions in the subsequent sections 
apply directly to D-branes of smaller dimension. For simplicity 
of the presentation, we shall stick to the case of a background
filling brane. 

\section{Perturbation Series for the Deformed Product} 

To obtain an explicit formula for the product $\star$ defined by 
rel. (\ref{ast0}), we regard the term $S_B$ (eq. (\ref{SB})) as 
a perturbation of the original bosonic theory with Neumann boundary 
conditions and study the usual field theoretic perturbation expansion. 
The correlators of the perturbed theory are constructed by the standard
prescription:
\ba
\langle \Phi_1 \dots \Phi_N\rangle^B_\e & = &
  \frac{1}{Z} \ \langle \Phi_1 \dots \Phi_N\  e^{-S_B} \rangle_\e
   \nn \\[2mm]
 & := &  \frac{1}{Z} \sum_{n=0}^{\infty} \left(\frac{-1}{4 \pi \ap}
  \right)^n
  \frac{1}{n!} \int_{\cH^\e_n} d^d z d^d\bz\  
 \langle \, \Phi_1 \dots \Phi_N \prod_{a=1}^n
   B_{i_aj_a} J^{i_a} (z_a)
  \bJ^{j_a}(\bz_a)\, \rangle\ \ .  \nn
\ea  
Here $Z := \langle \exp (-S_B)\rangle_\e $ and the expressions
depend on an UV-cutoff $\e$ through the domain of integration, 
$$ \cH^\e_n \ := \ \{\ (z_1,\dots,z_n)\ |\  \Im (z_a)\, >\, \e\ ,
\ |z_a - z_b| \, > \, \e \ \mbox{ for }\  a \neq b\ \} \ \ . $$
As long as $\e$ is positive, the integrals are protected against
divergencies caused by the short distance singularities of the
integrands \footnote{IR-divergencies can be cured by introducing 
a cutoff $\Lambda$ and restricting the integrations to $|z| < 
\Lambda$. We refrain from making this more explicit in our 
formulas.}.  When defining $\cH^\e_n$ we 
assumed that all fields $\Phi_\nu$ are inserted along the 
boundary. Eventually, we will choose $\Phi_1 = V[f](1)$ and 
$\Phi_2 = V[g](0)$.  
\smallskip
 
The correlators in the integrand can be evaluated with the help
of Ward identities for chiral currents. There exist two types
of terms in these Ward identities. One type involves contractions
between two currents while the second appears when we contract 
currents with one of the fields $\Phi_\nu$. Let us first look at 
contractions between two currents. Their contribution to the 
deformed correlation functions is evaluated with the help of the 
operator products expansions  
\be
    J^i(z) \ J^j(w) \ \sim \
    \frac{2 \ap G^{ij}}{(z-w)^2} \ + \ \dots\ \  
  \ \ , \ \ \ \ 
    J^i(z) \ \bJ^j(\bw) \ \sim \  \frac{2 \ap G^{ij} }{(z-\bw)^2} \
  \ + \ \dots \ \ . \label{JSB} 
\ee
Similar expansions exist for the $\bJ^i$ instead of $J^i$. 
We consider a perturbing operator inserted at the point 
$(z,\bz)$ and assume that its anti-holomorphic current $\bJ^j$ 
acts on one of the boundary fields while the holomorphic field 
$J^i$ is contracted with another current which may be either 
holomorphic or anti-holomorphic. In both cases, the 2-dimensional 
integral over the position $(z,\bz)$ of the insertion can be 
converted into a contour-integral. If we pair $J^i(z)$ with 
another holomorphic current, this contour-integral is easily 
seen to vanish. A non-trivial contribution arises only when 
we contract $J^i(z)$ with some anti-holomorphic current 
$\bJ^k(\bw)$. After the integration, the two perturbing fields 
at $(z,\bz)$ and $(w,\bw)$ turn out to be combined into only 
one insertion at $(w,\bw)$ which has the form: 
$$   \frac{1}{4 \pi \ap } \int_\cH dwd\bw
     \ \fri\  B_i^j \, B_{jl} \, J^i(w)\,  \bJ^l(\bw) \ \ .$$
In other words, the fields at $(z,\bz)$ have disappeared leaving 
an insertion at $(w,\bw)$ behind which is of the same form as the 
original expression for $S_B$ but with $\fri B^2$ appearing 
instead of $B$. Iteration of this argument allows us to sum over 
open chains of current contractions with arbitrary length 
\footnote{Note that closed loops of current-contractions 
are canceled by the denominator $Z$ of the deformed correlation
functions.}. With the correct combinatorial factors filled in, the 
terms form a geometric series. As a result, the perturbing fields 
are replaced by
\be S_B \ \rightarrow \  \frac{1}{4 \pi \ap } \int_\cH dzd\bz
     \ \left(\frac{1}{1+\i B}\right)_i^{\ l} \, B_{lj} \, 
      J^i(z)\,  \bJ^j(\bz) \ \ \label{help} \ee 
{\it and} we are no longer allowed to contract currents among each 
other. With this simple conclusion in mind we can turn towards 
the discussion of the second type of contractions in which 
currents act on one of the boundary fields. 
\smallskip

{}From now on, let us restrict our analysis to the computation of 
the operator product expansion (\ref{ast0}). Note that the leading 
contribution in the operator product expansion of open string vertex 
operators is completely determined by their 3-point functions. Hence, 
all our previous remarks on the perturbation of correlation functions 
apply directly to the perturbation of the operator product. In the 
following it is quite useful to decompose the matrix in the integral 
(\ref{help}) into its symmetric and anti-symmetric parts: 
\be 
     \frac{B}{1 + \i B} \ = \ \B^{\rm s} + \B^{\rm a} \ = \ 
     - \frac{\i B^2}{1+B^2} \ + \ \frac{B}{1+B^2}\ \ .       
\label{Bdec} 
\ee
The symmetric part $\B^{\rm s}$ gives rise to logarithmic divergencies
in the perturbation series for the operator product. They require  
renormalization of the boundary fields. We shall omit the detailed 
analysis at this point and only state the result: When the boundary 
fields $V[f](1)$ and $V[g](0)$ are properly renormalized, there remains 
no finite contribution coming from $\B^{\rm s}$, i.e.\ the symmetric 
matrix $\B^{\rm s}$ is  completely absorbed.  Hence, after renormalizing 
the boundary fields, we can work with the effective perturbing operator   
$$  S^{\rm eff}_B \ = \ \frac{1}{4 \pi \ap } \int_\cH dzd\bz
     \ \B_{ij} \, J^i(z)\,  \bJ^j(\bz) \ := \   
    \frac{1}{4 \pi \ap } \int_\cH dzd\bz
    \left(\frac{1}{1+ B^2}\right)_i^{\ l} \, B_{lj} \, 
      J^i(z)\,  \bJ^j(\bz) \ \  ,$$     
if we no longer allow for contractions between currents. Here and in 
the following we neglect to write the upper index $\ ^{\rm a}$ on the 
anti-symmetric matrix $\B = \B^a$.  
\smallskip

After all these preparations it is now straightforward to 
compute the deformed product $f \star g$ to all orders in 
perturbation theory: 
\ba
\left( V[f\, ] (1) \, V[g\, ](0) \right)^B & = &
  V[f\, ](1) \, V[g\, ](0)  
+ \frac{\ap}{\pi} \int_\cH dzd\bz
  \frac{1}{z-1}\frac{1}{\bz} \ \B^{ij} \
  V[\pl_i f\, ](1) \, V[\pl_j g\, ] (0) \nn \\[2mm]
& & \hspace*{1.1cm} + \frac{\ap}{\pi} \int_\cH dzd\bz
  \ \frac{1}{\bz-1}\frac{1}{z} \ \B^{ij} \
  V[\pl_j f\, ](1) \, V[\pl_i g\, ] (0) + O((\ap)^2) \nn \\[4mm]
 & = &
  V[f\, ](1) \, V[g\, ](0)  
+ \frac{\ap}{\pi} \int_\cH dzd\bz
  \left( \frac{1}{z-1}\frac{1}{\bz} -
  \frac{1}{\bz-1}\frac{1}{z}\right) \nn \\[2mm]
   & & \hspace*{4.7cm} \ \B^{ij} \
  V[\pl_i f\, ](1) \, V[\pl_j g\, ] (0) + O((\ap)^2) \nn \\[4mm]
& = &
 \sum_n \left(\frac{\ap}{\pi}\right)^n \frac{1}{n !}
\int dz_1 d\bz_1 \dots dz_n d\bz_n \prod_{a=1}^{n}
  \left( \frac{1}{z_a-1}\frac{1}{\bz_a} -
  \frac{1}{\bz_a-1}\frac{1}{z_a}\right)\nn \\[2mm]
  & & \hspace*{2.1cm}  
  \B^{i_1 j_1} \dots \B^{i_n j_n}
  V[\pl_{i_1} \dots \pl_{i_n} f\, ] (1)\ 
  V[\pl_{j_1} \dots \pl_{j_n} g\, ] (0)  \nn
  \\[4mm] 
& \sim &  \sum_n \left(\frac{\ap}{\pi}\right)^n \frac{1}{n !}
\int dz_1 d\bz_1 \dots dz_n d\bz_n \prod_{a=1}^{n}
  \left( \frac{1}{z_a-1}\frac{1}{\bz_a} -
  \frac{1}{\bz_a-1}\frac{1}{z_a}\right)\nn \\[2mm]
  & & \hspace*{1.8cm}  
  \B^{i_1 j_1} \dots \B^{i_n j_n}
  V[\pl_{i_1} \dots \pl_{i_n} f \ \pl_{j_1} \dots
   \pl_{j_n} g\, ] (0) \ + \ \dots \  \ . \nn
\ea
In the last step we have kept only the leading
contribution from the operator product expansion for
vertex operators at $B = 0$. All other manipulations were
exact. To understand the expression for the $n^{\rm th}$
summand, it suffices to look at the first order terms.
Note that the second term on the right hand side of
the first line is connected with the action of $J^i$ on
the open string vertex operator at the point $x=1$ and
of $\bJ^j$ on the vertex operator at $x=0$. For the second
line one needs to interchange the role of $J^i$ and $\bJ^j$.
There are certainly terms where both currents act on the
same vertex operator. These terms vanish since they involve
a contraction of $\pl_i \pl_j$ with the antisymmetric matrix
$\B^{ij}$.   
\smallskip

We could certainly compute the remaining integrals in the 
previous expression for operator product expansion. But we 
shall leave them in their present form and produce a more 
compact answer by introducing the following shorthand 
notations: 
\ba & & w_n \ := \  \frac{1}{(2\pi)^{2n}} \frac{1}{n !}
\int d^n z d^n\bz \prod_{a=1}^{n}
  \left( \frac{1}{z_a-1}\frac{1}{\bz_a} -
  \frac{1}{\bz_a-1}\frac{1}{z_a}\right) \ \ , \nn \\[3mm]
  & & \bB_n(f,g) \ := \ \sum  \B^{i_1 j_1} \dots \B^{i_n j_n}
   \ \pl_{i_1} \dots \pl_{i_n} f \ \pl_{j_1} \dots
   \pl_{j_n} g   \ \ . \nn 
\ea
Our perturbative construction of the operator product for
open string vertex operators provides us with an explicit
formula for the product $f \star g$ defined through eq.
(\ref{ast0}). It is given by
\be 
 f \star g \ = \ \sum_n \, (4 \pi \ap)^n\, w_n\,  \bB_n(f,g) \ \ .
\label{ast1} \ee
The remarks above guarantee that $\star$ is associative
but one may also check this directly now. On the other
hand, it is not commutative as we infer e.g.\ from
the first order term in the commutator:
$$ f \star g - g \star f \ = \  \kappa \, \ap \, \B^{ij}\
    \pl_i f \ \pl_j g \ + \  O((\ap)^2) \ = \ \kappa\, \ap
    \, \{\, f\, ,\, g\, \}_\B \ + \ O((\ap)^2)\ \ , $$
where $\kappa$ is some constant factor and we have
introduced the Poisson bracket $\{\,.\, ,\, .\, \}_\B$
through the second equality. The expression (\ref{ast1}) 
for $\star$ coincides with Kontsevich's formula for the 
Moyal product of functions on the background $\R^d$ \cite{Kon}.
{}From this we conclude that operator products of boundary fields 
give rise to the usual Moyal deformation in the direction of the 
`Poisson bi-vector' 
\be \label{al} \B \ = \ B \ (1+ B^2)^{-1} \ \ . \ee
Note that $\B \sim B$ for small magnetic fields while 
$\B \sim B^{-1}$ when $B^2$ becomes very large. We shall 
refer to the region of very strong fields as `topological 
regime'. In this limit, the metric $G$ can be neglected in 
comparison to the B-field. Our results for strong fields 
are consistent with the recent analysis \cite{CaFe} of 
topological $\sigma$-models which predicts the deformation 
for large $B^2$ to be in the direction of $B^{-1}$. The 
latter is the `Poisson bi-vector' associated with the 
`symplectic  form' $B$. Let us also remark that the 
expression (\ref{al}) is known from the theory of toroidal 
compactifications where is appears as the B-field on the 
dual torus. 
\medskip

Before we conclude this section, we would like to discuss
briefly how the boundary currents $J^i(x) = \bar J^i(x)$ of the 
original field theory make their appearance in the world-volume 
geometry. With our previous experience on the perturbative 
computation of operator products, it is rather easy to 
see that
$$ \left( J^i(1) \ V[f](0) \right)^B \ \sim \ 
     \frac{2 \ap}{\i} \, G^{ij}\,  
      \left(\frac{1}{1+B^2}\right)_j^k 
     \ V[\partial_k f](0) \ + \ \dots \ \ . $$
Hence, the chiral boundary currents describe infinitesimal symmetries 
of the D-brane, i.e.\ its world-volume algebra inherits derivations 
$\delta^i$ of the form 
\be \delta^i\ = \  G^{ij}\,  
      \left((1+B^2)^{-1}\right)_j^k \ \partial_k  
 \label{deriv} \ \ . 
\ee
In the limit of vanishing B-field, $\delta^i$ coincides with 
the usual derivative $G^{ij}\partial_j$. As one would expect, 
the number of infinitesimal translation symmetries of a D-brane
does not depend on the field strength as long as the background
field is constant. 

\section{Extension to Fermionic Fields} 

\def\bpsi{\bar \psi} 
\def\bpl{\bar \pl} 
In this final section we would like to extend our analysis to 
fermionic fields. We consider a $d$-plet of Majorana fermions 
described in terms of the usual (anti-)holomorphic components 
$\psi^i(z), \bpsi^i(\bz)$. There exists a choice of considering
the fermionic fields in the Ramond or the Neveu-Schwarz sector. 
Here we are only interested in the latter. The Ramond sector 
turns out to provide a module of the non-commutative algebra 
that we shall obtain from the Neveu-Schwarz sector.  
\smallskip

Recall that the operator product expansions for the fermionic 
fields are of the form:
$$ \psi^i(z)\  \psi^j(w) \ \sim \ \frac{G^{ij}}{z-w}
  \ + \ \dots \ \ \ \ \ , \ \ \ \ \    
\psi^i(z)\ \bpsi^j(\bw) \ \sim \ \frac{G^{ij}}{z-\bw}
\ + \ \dots $$ 
and similar expressions hold when holomorphic fields are replaced
by  anti-holomorphic ones (and vice versa). This motivates to 
assign the fermionic fields $\psi^i(x) = V[\eta^i](x)$ to 
generators $\eta^i$ of a Clifford algebra with the 
multiplication 
$$ [\, \eta^i\, , \, \eta^j \, ]_+ \ = \ G^{ij}\ \ . $$
In this way we may re-express the operator product expansions
of fermionic boundary fields in the form $V[\eta^i](1) V[\eta^j](0) 
= V[[\eta^i,\eta^j]_+](0) = V[\,G^{ij}](0) = G^{ij}$.   
\smallskip

Placing this theory of $d$ Majorana fermions in a non-vanishing 
B-field means to perturb the original theory by the term 
$$ S'_B \ = \ \frac{1}{8 \pi } \int_\cH dz d\bz\ B_{ij} 
\left( \psi^i(z)\, \pl \psi^j(z) - \bpsi^i(\bz)\, \pl \bpsi^j(\bz) 
  \right) \ \ . $$
When combined with the bosonic sector, the contribution $S'_B$ ensures 
supersummetry of the total theory with boundary conditions $\psi_-^i = 
\i B^i_j \psi^j_+$ in the fermionic sector. Here we have introduced 
$\psi^k_\pm = \psi^k \pm \bpsi^k$. 
\smallskip
 
Our evaluation of the perturbation expansion proceeds very 
much as in the case of bosonic fields. Once more, contributions 
{}from chains of contractions between fields in the perturbing 
operator can be summed up. This is achieved my means of the 
standard formula 
$$  \int \i dz d\bz \ f(z)\ \bpl_z \frac{1}{z-w} \ = \  2 \pi 
    \, f(w) \ \  $$ 
and a similar expression with the role of holomorphic and 
anti-holomorphic variables interchanged. The procedure allows 
us to replace the B-field in $S'_B$ by $B\, (1 + iB)^{-1}$, if 
at the same time we refrain from further contractions between 
the effective insertions. 
\smallskip
    
The aim now is to compute the leading term in the operator
product expansion of $\psi^i(1)$ and $\psi^j(0)$. To evaluate 
the contractions between the `effective' perturbing field and
the boundary operators we insert the decomposition (\ref{Bdec})
of $B(1+\i B)^{-1}$ into its symmetric and anti-symmetric parts. 
Because of the anti-commutativity of the fermionic fields, only 
the anti-symmetric matrix $\B^a$ enters into the final formula 
for the deformed operator product:   
$$\left( \psi^i(1)\ \psi^j(0)\right)^B \ \sim \ G^{ij} + G^{ik} \left( 
  \frac{-B^2}{1+B^2}\right)_k^j \ + \ \dots \ = \  
  G^{ik}\left( \frac{1}{1+B^2} \right)_k^j\ + \ \dots \ \ . $$
We can translate this result into a deformation of the Clifford
algebra. Under the influence of the B-field, the original 
generators $\eta^i$ obey the deformed relations  
\be [\ \eta^i\, \stackrel{\star}{,} \, \eta^j \ ]_+ \ = \ G^{ik}
   \left( \frac{1}{1+B^2} \right)_k^j \ \ . 
\label{Cdef} \ee
The combination on the right hand side appeared already in eq.\ 
(\ref{deriv}) for the derivations of the world-volume algebra. 
\smallskip

The model we have studied possesses an $N=1$ supersymmetry on the 
world-sheet but our considerations certainly generalize to theories 
with more supersymmetries. As in the case of closed strings \cite{SNCG}, 
the non-commutative world-volume geometries inherit Dirac operators from 
the supersymmetry generators of the field theory. In the context of 
non-commutative geometry \cite{ConB}, they give rise to differential 
geometries on the branes' world-volumes.

\section{Conclusions} 

In this text we have constructed the world-volume geometry (eqs. 
(\ref{ast1}),(\ref{Cdef})) of D-branes in flat backgrounds. 
Note, however, that both the underlying concepts and the proposed 
techniques can be applied beyond these simple examples. In 
particular, it is possible to study the perturbative expansion 
for the operator product of open string vertex operators in 
non-linear $\sigma$-models. To treat the dependence of B-fields
on co-ordinates of the background one makes use of standard 
background field methods \cite{AFM}. It is quite remarkable that 
the resulting expansions are still organized very much as in the 
corresponding version of Kontsevich's quantization formula. We 
shall report on these issues in a forthcoming paper.

More concretely, one may try to reconstruct the world-volume 
of D-branes for some specific (CFT-) backgrounds, such as e.g.\ 
the WZW-model. The classical world-volume of branes on group 
manifolds is given by certain `integer' conjugacy classes of 
the group \cite{WZW1} and the branes come equipped with a 
B-field. In case of the SU(2)-WZW theory, their world-volumes are 
described by fuzzy spheres, at least in an appropriate limit 
\cite{WZW2}.
\medskip

It might also be interesting to study effective field theories 
on general D-brane world-volumes within the presented framework. 
The construction of the effective actions is a problem in boundary 
perturbation theory. Since the perturbing boundary operators and 
the `generators' of the world-volume algebra appear on equal 
footing, it should be possible to develop a rather general approach 
towards effective theories on non-commutative spaces. 

The techniques of \cite{ReSc} provide another step in this direction. 
It was shown there that non-commutativity of the boundary operator
product expansion is the only obstruction in boundary deformation 
theory. In this sense, a detailed knowledge about operator products
of boundary fields (and about their non-commutativity) is essential 
for our understanding of general boundary flows and of D-brane moduli 
spaces, in particular.    
\bigskip
\bigskip
\bigskip

\noindent
{\bf Acknowledgements:} I would like to thank
A.Yu.\ Alekseev, R.\ Dijkgraaf, A.\ Konechny, 
N.P.\ Landsman, E.\ Langman, A.\ Recknagel, A.\ 
Schwarz and S.\ Theisen for stimulating 
discussions and remarks.

\end{document}